\documentclass[11pt,english]{article}
\usepackage{lmodern}

\usepackage[T1]{fontenc}
\usepackage[latin9]{inputenc}
\usepackage[letterpaper]{geometry}
\usepackage{babel}
\usepackage{amsthm}
\usepackage{amsmath}
\usepackage{amssymb}
\usepackage{graphicx}
\usepackage[unicode=true,pdfusetitle,
 bookmarks=true,bookmarksnumbered=false,bookmarksopen=false,
 breaklinks=false,pdfborder={0 0 1},backref=false,colorlinks=false]
 {hyperref}

\makeatletter

\providecommand{\tabularnewline}{\\}

\numberwithin{equation}{section}
\numberwithin{figure}{section}
\numberwithin{table}{section}
\newcommand{\lyxaddress}[1]{
\par {\raggedright #1
\vspace{1.4em}
\noindent\par}
}

\makeatother

\begin{document}

\title{Intermittency at Fine Scales and Complex Singularities of Turbulent
Couette Flow}

\date{\vspace{-0.5cm}
}

\author{Andre Souza and Divakar Viswanath }

\maketitle

\lyxaddress{Department of Mathematics, University of Michigan (sandre/divakar@umich.edu). }
\begin{abstract}
Fine scales of turbulent velocity fields, beyond the inertial range
and well into the dissipative range, are highly intermittent. It has
been hypothesized that complex plane singularities are the principal
mechanism behind fine scale intermittency. In this article, we view
the velocity field of a turbulent flow as an analytic function of
time. Although the function is only available for real values of time,
we present a numerical technique to analytically continue the function
to complex values of time, and with sufficient fidelity to locate
and visualize the singularity closest to the real axis. Using this
technique, we demonstrate a robust connection between temporal intermittency
and the location of singularities in the complex plane.
\end{abstract}

\section{Introduction}

Intermittency in turbulent flows refers to sharp and spatially isolated
peaks of energy contained in large wave numbers, beyond the inertial
range and well into the dissipative range. In the words of Batchelor
and Townsend \cite{BatchelorTownsend1949}, who discovered this phenomenon,
``the energy associated with large wave-numbers is very unevenly
distributed in space'' and ``there appear to be isolated regions
in which the large wave-numbers are 'activated', separated by regions
of comparative quiescence.'' This fine scale intermittency is a fundamental
feature of turbulent flows \cite{Batchelor1953,Frisch1995,Sreenivasan1991}. 

Frisch and Morf \cite{FrischMorf1981} hypothesized that intermittency
is a manifestation of singularities in the complex plane. Thus the
activation mechanism is believed to be the occurrence of singularities
in the complex plane, with the intermittent peak occurring right below
the singularity closest to the real axis. Complex singularities are
intrinsically easier to study with a single independent variable.
Therefore Frisch and Morf shifted the emphasis to temporal intermittency,
with time as the single independent variable with the entire velocity
field viewed as a function of time.

The connection between intermittency and complex singularities has
been demonstrated in a variety of systems such as Langevin's and Burger's
equations \cite{FrischMorf1981,PaulsFrisch2007}. However, to the
best of the authors' knowledge, such a connection is yet to be demonstrated
for the incompressible Navier-Stokes equations in the turbulent regime.
In this article, we develop a numerical technique that is applicable
to trajectories of large scale PDE. Using that technique, we demonstrate
a robust connection between temporal intermittency and complex singularities.

Here at the outset, we reprise a beautiful argument of Kraichnan \cite{Kraichnan1967}.
The argument is heuristic but gives powerful intuition regarding the
phenomenon of fine scale intermittency, partially explaining many
of the plots found later in this article and showing why the manifestation
of intermittency is clearest in the dissipative range. The argument
is as follows. Suppose that the velocity field of the turbulent flow
is infinitely differentiable with each derivative being square integrable.
Direct numerical simulation of turbulence from the incompressible
Navier-Stokes equations implicitly assumes the velocity field to be
analytic, and the great success of direct numerical simulation is
certainly evidence for the validity of that assumption. The assumption
of smoothness implies that energy should fall off faster than algebraically
(and exponentially fast if the velocity field is analytic) with increase
in wave-number. The super-algebraic and possibly exponential decay
of energy with wave-number means that a change in wave-number ``by
a few percent'' implies an ``enormous'' change in the energy associated
with the wave-number. If the energy spectra is governed by a length
scale that varies gently from region to region, this gentle variation
is enormously amplified and shows up as intermittency in the high
wave-numbers. In essence, the hypothesis of Frisch and Morf \cite{FrischMorf1981},
investigated and validated in this article, is that the gentle variation
in the large which gets amplified into intermittency in fine scales
is due to singularities in the complex plane. 

Energy dissipation per unit mass, denoted $\epsilon$, is an important
parameter in turbulence. Indeed, it is the \emph{only} parameter in
the Kolmogorov theory of the inertial range \cite{Batchelor1953}.
Fine scale intermittency may be detected experimentally through an
analysis of the variation in $\epsilon$ \cite{MeneveauSreenivasan1987,MeneveauSreenivasan1991}.
Thus physical space analysis of turbulent velocity fields obtained
from numerical simulation suffices for demonstration of spatial intermittency
as well as for isolating structures associated with that phenomenon
\cite{SchumacherSreenivasanYeung2005}.

The tack we take in this article is different, and possibly complementary.
We study fine scale intermittency through analytic continuation into
the complex $t$-plane. An advantage is a direct connection to the
hypothesized mechanism. Continuation into the complex $t$-plane could
be complementary to physical space analysis as we discuss in the concluding
section. 

Analytic continuation into the complex plane is infamous for its numerical
difficulty. The technique of Padé approximation, and strategies for
better numerical stability, have been set forth by Weideman \cite{Weideman2003},
Gonnet, Pachón, and Trefethen \cite{GonnetPachonTrefethen2011} and
Webb \cite{Webb2013}. A crucial ingredient is the use of complex
phase plots championed by Wegert and others \cite{Wegert2012,WegertSemmler2010},
and whose pertinence to the computation of complex singularities was
communicated to us in person by Hrothgar and Trefethen. These phase
plots play a central role in our technique and we rely on them to
locate singularities in the complex plane. The phase plots proved
to be a more viable approach to stabilize numerical computations than
strategies for deflating tiny and repeated singular values developed
in \cite{GonnetPachonTrefethen2011}. 

All numerical simulations are carried out with constant time steps
for reasons explained in section 2. The numerical method for locating
complex singularities described in section 2 interpolates to Chebyshev
points of the second kind from equi-spaced data. The Chebyshev series
are filtered before computing Padé approximants. Many of the ingredients
in the numerical method we describe have occurred in earlier work
in some form. Our contribution is to synthesize a numerical method
that works robustly for turbulent flows. 

Somewhat counter-intuitively, the high dimensionality of discretizations
of turbulent velocity fields appears to be a help rather than a hindrance.
The numerical method of section 2 takes advantage of intermittency
in high wave-numbers in locating the complex singularities. 

In section 3, we compute the complex singularities of six periodic
or relative periodic solutions of plane Couette flow. The plane Couette
flow set-up we use is the minimal flow unit \cite{HamiltonKimWaleffe1995}.
The minimal flow unit was derived by constraining turbulent Couette
flow into a small box at the low Reynolds number of $Re=400$. Since
the Reynolds number is low, the dissipative range is reached easily.
Another advantage of the minimal flow unit is its connection to the
self-sustaining process developed by Waleffe \cite{Waleffe1997}.
While inertial range turbulence is statistical, turbulence is also
characterized by coherent structures. The self-sustaining process
emphasizes the dynamical aspects of fine-scale turbulence with connections
to coherent structures. 

The six periodic and relative periodic orbits, whose complex singularities
are computed, are from earlier work \cite{Viswanath2007} . For each
of these orbits, we demonstrate a clear connection between intermittency
and the location of complex singularities. There are no profound advantages
to using periodic orbits. One could use any turbulent trajectory.
But as a practical matter there are two reasons for opting for these
solutions of plane Couette flow. Firstly, the orbits are reported
in \cite{Viswanath2007} with estimates for their precision so that
they may be reproduced. Secondly, the earlier study of these orbits
has identified different regimes enabling a preliminary investigation
of the connection between observed turbulence and the location of
complex singularities.

The concluding section 4 discusses further connections of the work
presented in this article. One of these is to the spatial analysis
of fine scale intermittency, already alluded to above. Another is
to blow-ups in physical space.

\section{Numerical method}

The numerical method used for computing singularities will be described
using the Lorenz example. The Lorenz equations $\dot{x}=10(y-x),\,\dot{y}=28x-y-xz,\:\dot{z}=-8z/3+xy$
admit singular solutions that may be expanded in convergent psi-series
as follows \cite{ViswanathSahutoglu2010}:
\begin{alignat}{2}
x(t) & = & \frac{P_{-1}(\eta)}{t-t_{0}}+P_{0}(\eta)+P_{1}(\eta)(t-t_{0})+P_{2}(\eta)(t-t_{0})^{2}+\cdots\nonumber \\
y(t) & =\frac{Q_{-2}(\eta)}{(t-t_{0})^{2}}+ & \frac{Q_{-1}(\eta)}{t-t_{0}}+Q_{0}(\eta)+Q_{1}(\eta)(t-t_{0})+Q_{2}(\eta)(t-t_{0})^{2}+\cdots\nonumber \\
z(t) & =\frac{R_{-2}(\eta)}{(t-t_{0})^{2}}+ & \frac{R_{-1}(\eta)}{t-t_{0}}+R_{0}(\eta)+R_{1}(\eta)(t-t_{0})+R_{2}(\eta)(t-t_{0})^{2}+\cdots\label{eq:lrz-psi-series}
\end{alignat}
Here $\eta=\log(t-t_{0})$ with the branch cut chosen appropriately,
and $P_{i},Q_{i},R_{i}$ are polynomials in $\eta$. These polynomials
are given explicitly in earlier work \cite{ViswanathSahutoglu2010}.
Here we note that $Q_{-2}=-i/5$, $R_{-2}=1/5$, $P_{-1}=Q_{-1}=2i$,
and $R_{-1}=17/9$, so that the dominant part of the singularity is
a pole for $x(t)$ and a double pole for $y(t)$ and $z(t)$. The
first terms in the psi-series which are not constants are 
\[
Q_{-1}=-{\frac{349}{81}}\, i-{\frac{988}{81}}\, i(\eta+C),\:\: R_{-1}={\frac{1385}{54}}-{\frac{988}{81}}\,(\eta+C),
\]
where $C$ is an undetermined constant. There is another undetermined
constant $D$ in the psi-series which occurs for the first time in
$Q_{2}$.

Not all the singularities of Lorenz are proved to be psi-series of
this form. However, a few singularities were examined numerically
in \cite{ViswanathSahutoglu2010} and were all found to be psi-series
of the form given above. In addition, the locations of the singularities
were determined with $10$ digits of precision, which makes the Lorenz
example useful for validating the numerical method described here.

The Lorenz example, being of only $3$ dimensions, can be tackled
using a number of techniques. It is amenable to extended precision
computations. Orbits of the Lorenz system were computed with $500$
digits of accuracy in \cite{ViswanathSahutoglu2010}. For the incompressible
Navier-Stokes equations in the fully turbulent regime, simply integrating
the trajectories is among the most complex and difficult computations.
The room for deploying techniques that find complex plane singularities
is much more limited. 

In this section, we refine Padé based techniques \cite{GonnetPachonTrefethen2011,Weideman2003}
and derive a numerical method that is reliable when applied to turbulent
signals. The numerical method is described in three stages. The first
stage is an analysis in the purely real time domain. The second stage
is the computation of a Padé approximant. The third stage is a discussion
of the Wegert--style phase plots of the Padé approximant. Complex
singularities are located using phase plots, and we compare the locations
obtained through the phase plots to the far more accurate computations
of \cite{ViswanathSahutoglu2010}.

\subsection{Chebyshev series and its analysis}

To describe the numerical method, we look at the $z$-coordinate of
the $AABB$ orbit of the Lorenz system. For the nomenclature of Lorenz
orbits, see \cite{Viswanath2003}. The $AABB$ orbit has period equal
to $3.084276...$ 

The Lorenz orbits can be computed using a special method described
in \cite{Viswanath2003}. Turbulence trajectories are obtained using
ordinary direct numerical simulation, and to obtain a reliable comparison
the Lorenz orbits too were computed using numerical integration. There
is one important point pertinent to direct numerical simulation, however.
The time-steps must be constant. 

If a Hamiltonian system is integrated using constant time-steps and
a symplectic discretization, it has been proved that the numerical
trajectory (ignoring rounding errors) approximates the trajectory
of a perturbed Hamiltonian with error that is exponentially small
in the time-step for an interval of time that is exponentially long
in the time-step \cite{HairerLubich1997}. The arguments of \cite{HairerLubich1997}
can be made to apply to non-Hamiltonian ordinary differential equations
without major modifications. Thus if the time-step is constant, there
is reason to think that the effect of the discretization error is
to introduce a smooth and analytic perturbation of the underlying
differential equation. On the other hand, adaptive time-stepping strategies
are non-smooth, and even though they allow for longer time-steps,
they destroy the analytic structure of the underlying differential
equation.

Once a signal $u(i\Delta t)$ is obtained for integer values of $i$,
the next step is to pick a time interval and compute a Chebyshev series.
Suppose that the time interval is $[a,b]$ with $a=i_{a}\Delta t$
and $b=i_{b}\Delta t$. To convert, the signal $u(t)$, $a\leq t\leq b$,
into a Chebyshev series, we first interpolate the signal at Chebyshev
points (of the second kind) $\cos(j\pi/n)$, $j=0,1,\ldots,n$, shifted
for the interval $[-1,1]$ to the interval $[a,b]$. If $t_{j}$ is
a shifted Chebyshev point in $[a,b]$ and $t_{j}\in[i\Delta t,(i+1)\Delta t]$,
the value of $u(t_{j})$ is computed through polynomial interpolation
using the nodes $t=\left(i-k\right)\Delta t,\,(i-k+1)\Delta t,\ldots,(i+k+1)\Delta t$. 

The polynomial interpolation is from equi-spaced data. However, there
is no need to fear the Runge phenomenon as $t_{j}$ is near the center
of the domain of interpolation. In addition, $k$ is small, with $k=4$
in almost all reported computations ($k=4$ implies at least $8$-th
order accuracy). The interpolation was carried using the barycentric
Lagrange formula, which is known to have excellent numerical stability
\cite{BerrutTrefethen2004}. Interpolated Chebyshev data was moved
back to the original equi-spaced grid to assess the quality of interpolation,
and to ensure that no information was lost during interpolation.

Once the signal $u(t)$ is available at Chebyshev points shifted to
the interval $[a,b]$, it is converted to a Chebyshev series using
the discrete cosine transform. If $T_{k}(x)=\frac{1}{2}\left(z^{k}+\frac{1}{z^{k}}\right)$
denotes the Chebyshev polynomial of degree $n$, with $z$ determined
from $x=\frac{1}{2}\left(z+\frac{1}{z}\right)$, and $\tilde{T}_{k}$
is $T_{k}$ shifted from the interval $[-1,1]$ to the interval $[a,b]$,
the Chebyshev series obtained from the discrete cosine transform is
of the form 
\[
u(t)=c_{0}+c_{1}\tilde{T}_{1}(t)+c_{2}\tilde{T}_{2}+\cdots+c_{n}\tilde{T}_{n}
\]
and is valid for $a\leq t\leq b$. This Chebyshev series will be converted
into a Padé approximant later. It is important to do a preliminary
analysis of the coefficients $c_{i}$ before computing the Padé approximant.
Much of the analysis is visual as we will now explain.

\begin{figure}

\includegraphics[width=2in,height=2in]{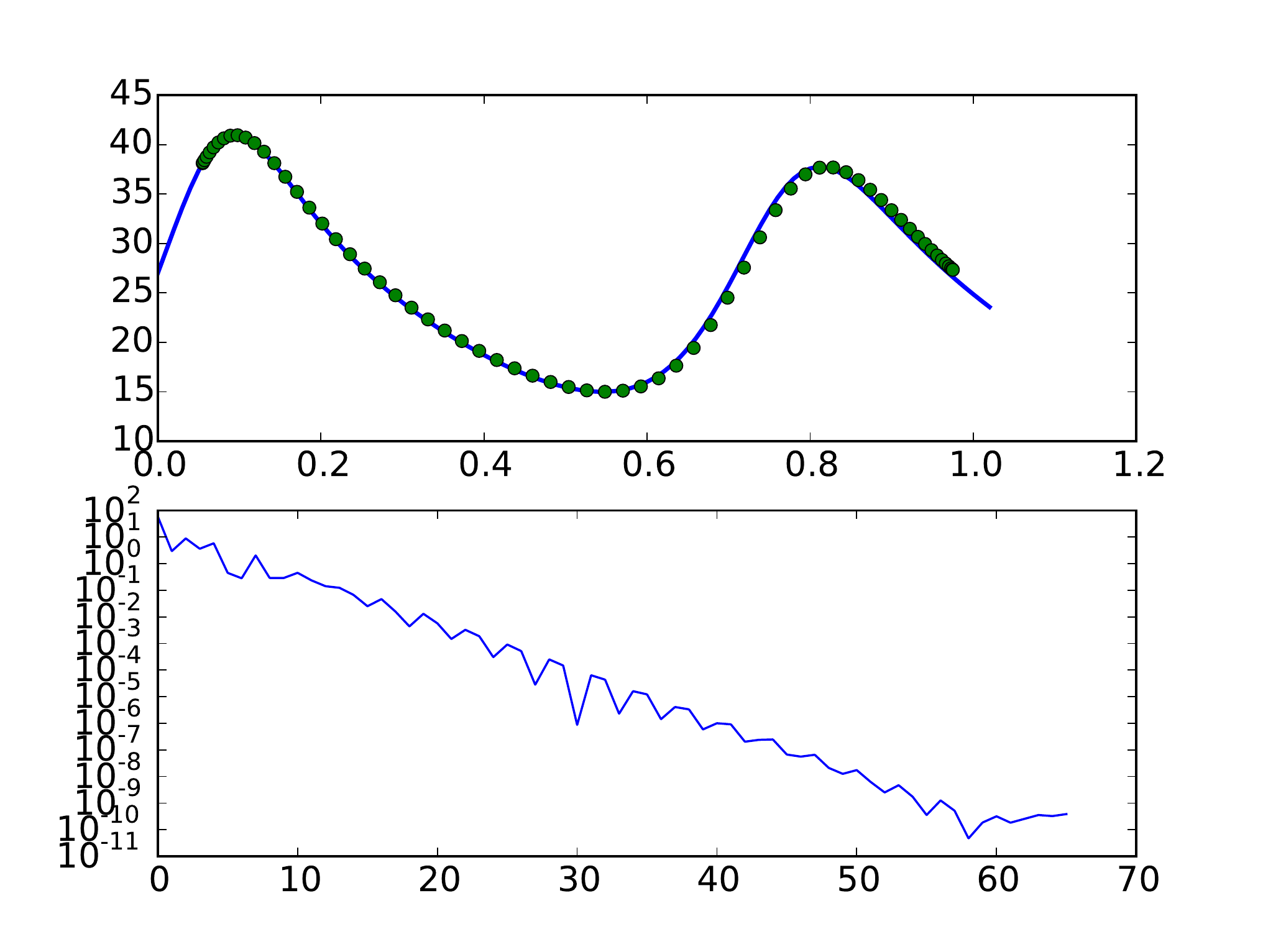}\includegraphics[width=2in,height=2in]{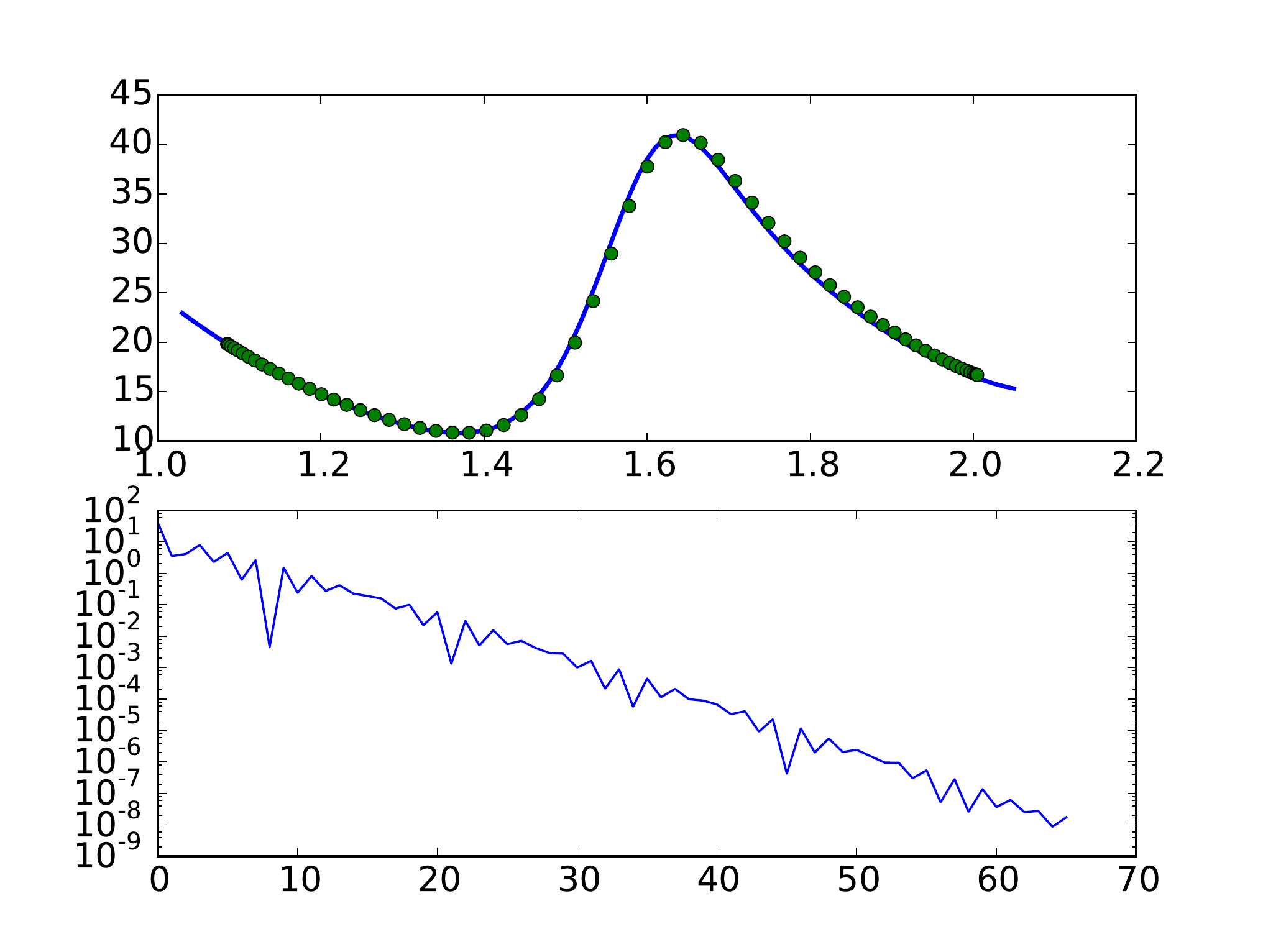}\includegraphics[width=2in,height=2in]{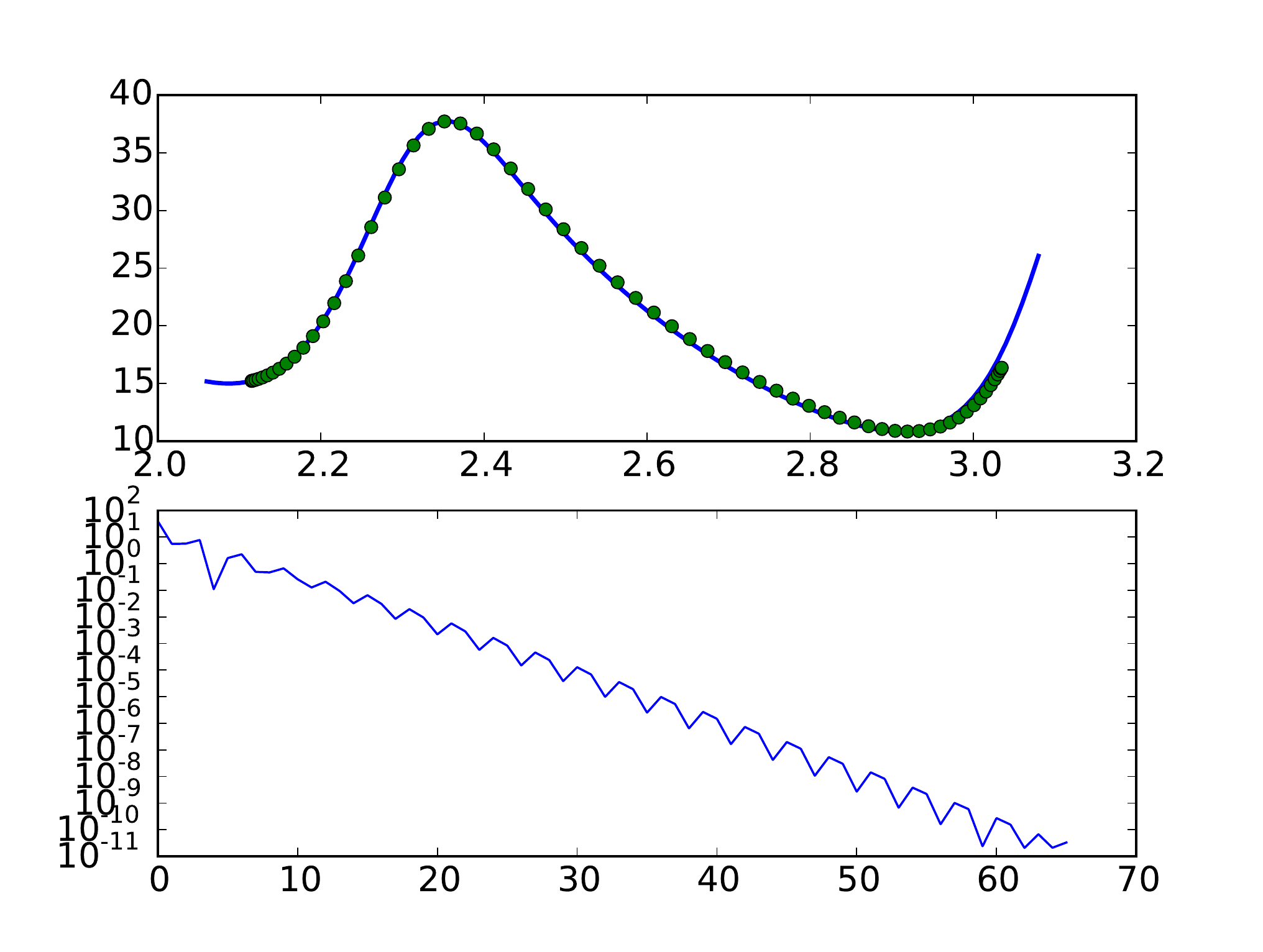}

\caption{Plots of the $z$-coordinate of the $AABB$ Lorenz orbit. The top
plots show $z$ as a function of time, with the dots being interpolants
at Chebyshev points. The bottom plots are $|c_{n}|$, absolute values
of coefficients of the Chebyshev series, of the top plots graphed
against $n$.\label{fig:lrz-seg-plots}}

\end{figure}

Figure \ref{fig:lrz-seg-plots} shows plots for the Lorenz orbit $AABB$.
The period of the orbit has been split into three equal segments.
It is easily noticeable that the Chebyshev series of the middle segment
is cleaner with a more pronounced pattern. The Chebyshev coefficients
decrease exponentially all three segments, but in the middle segment
we see spikes that protrude down with a clear rhythm. Isolating a
Chebyshev series with this kind of pattern is crucial. The Padé approximant
relies on this kind of a pattern to locate the complex singularity.
If the pattern is not clear to begin with the Padé approximant will
not work too well. 

For some of the Lorenz orbits and even turbulent orbits of plane Couette
flow, we are able to find Chebyshev series that exhibit patterns that
are even cleaner and more pronounced than the middle segment in Figure
\ref{fig:lrz-seg-plots}. For others, the pattern is not as clean.
In addition to the pattern being clean and well-established, it is
advantageous if the pattern starts early. 

\begin{figure}

\includegraphics[scale=0.4]{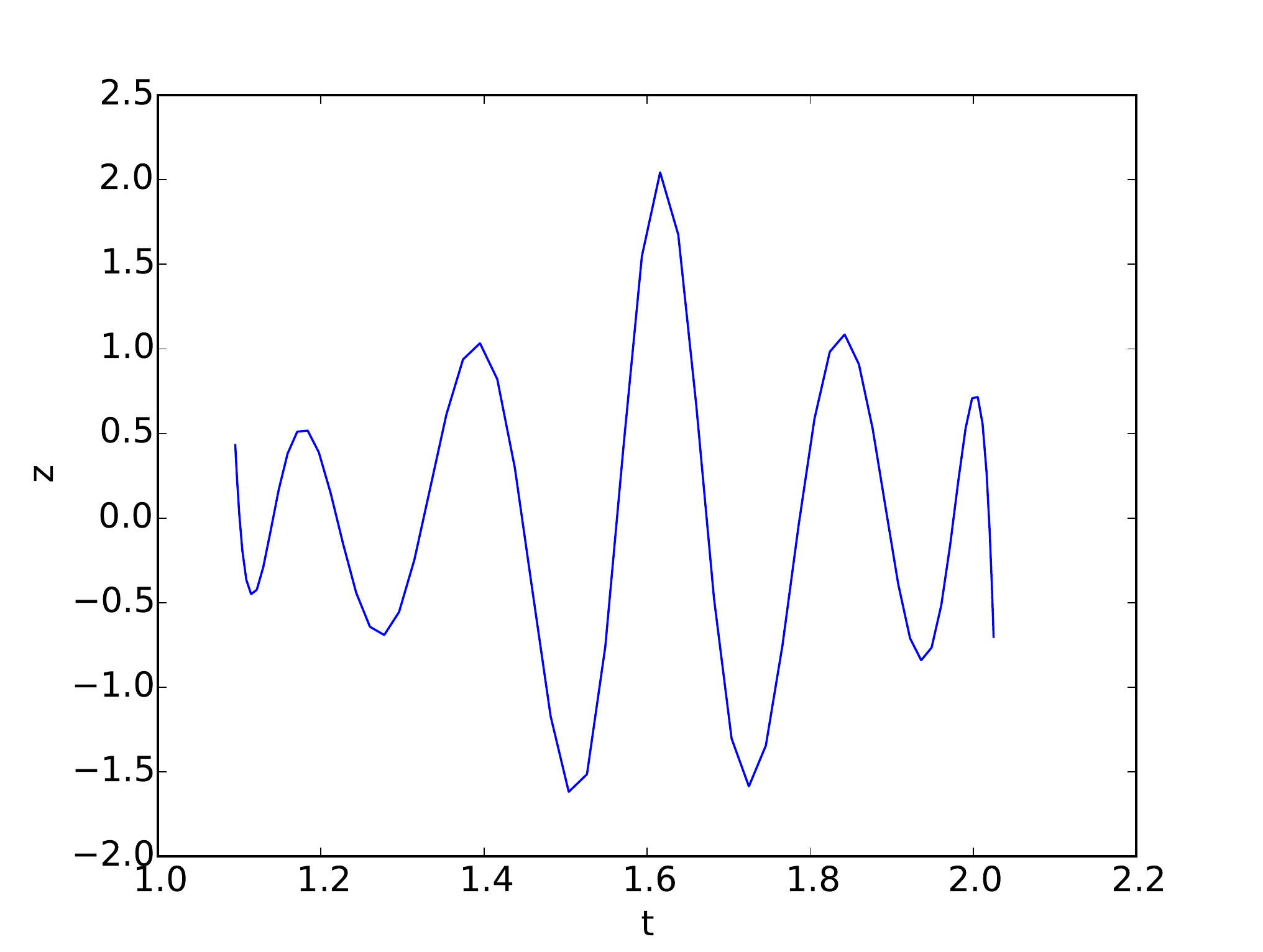}\includegraphics[scale=0.4]{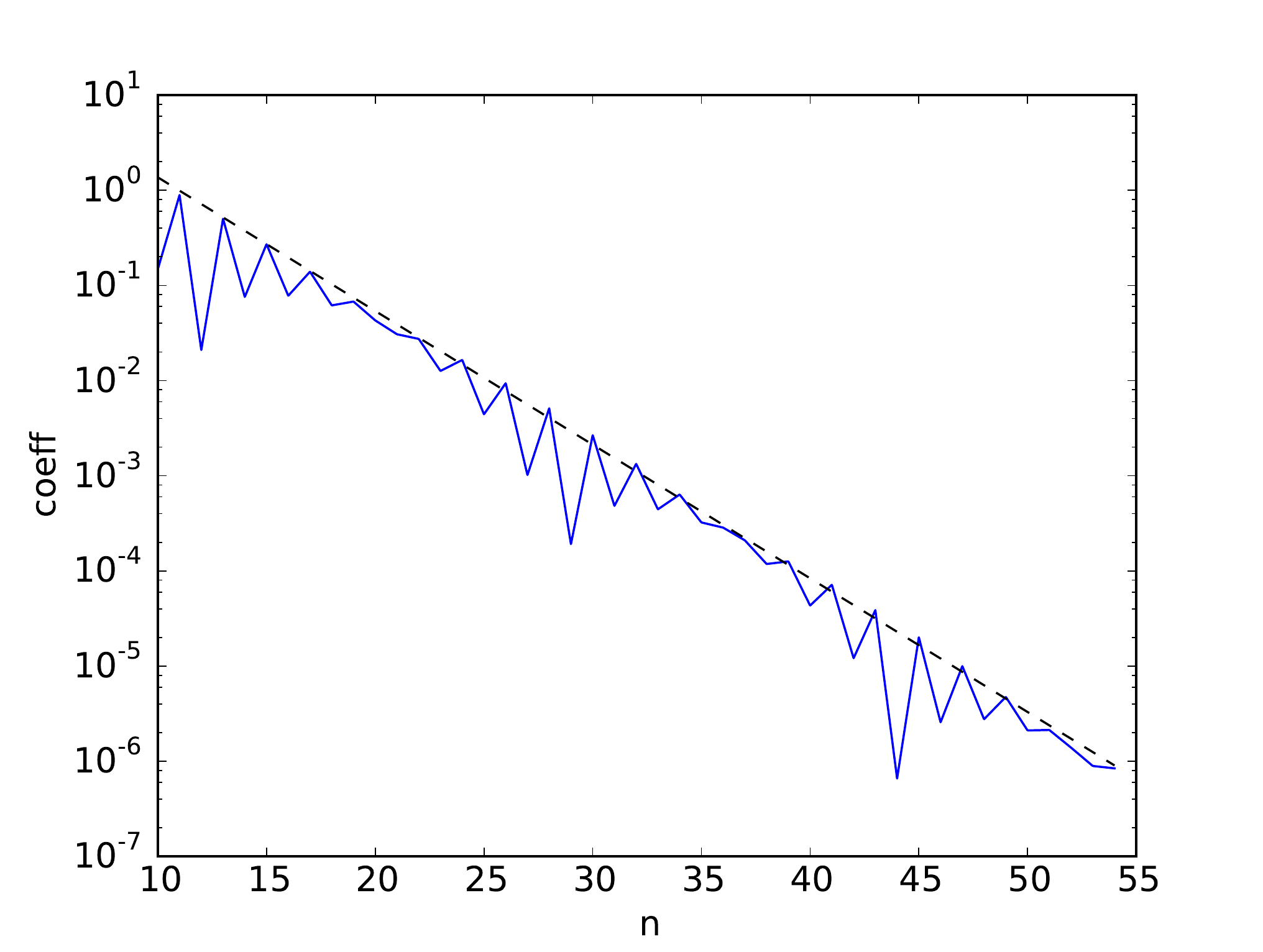}\caption{Filtered signal corresponding to the middle segment of Figure \ref{fig:lrz-seg-plots}
and a linear fit of its Chebyshev coefficients.\label{fig:lrz-filtering}}

\end{figure}

Once a Chebyshev series with a clean pattern has been isolated, the
pattern can be improved further using filtering as shown in Figure
\ref{fig:lrz-filtering}. Here the Chebyshev series of the middle
segment was filtered by setting to zero the first $10$ coefficients,
since they are not part of a pattern, as well as the last few which
were deemed to be not sufficiently accurate. The filtered signal on
the left side of Figure \ref{fig:lrz-filtering} almost gives away
the real part of the location of the singularity as $\Re(t)\approx1.6$.
Filtering makes the intermittency more pronounced.

Identification of the role of filtering in accentuating intermittency
is due to Frisch and co-workers \cite{FrischMorf1981,PaulsFrisch2007}.
As noted by those authors, the roots of the idea are found in the
theory of complex variables as developed in the late 1800s and early
1900s. In that body of work, the asymptotics of the coefficients of
a power series are connected to the form and location of the singularities.
The further out one goes into the series, the clearer the asymptotics
will be. In numerical work, however, one cannot go too far because
of accuracy issues.

Alternatively, the role of filtering may be understood as follows.
Suppose that $f(t)$ is analytic except for singularities at $t=a\pm ib$,
with $b\neq0$, in the complex plane. Let $t_{0}\in\mathbb{R}$. Consider
the expansion of $f(t)$ in powers of $(t-t_{0})$ with $t_{0}$ varying
along the real line. The radius of convergence is least at $t_{0}=a$
and the coefficient of $(t-t_{0})^{n}$ spikes at $t_{0}=a$, for
large $n$. The increase in steepness of the spike at $t_{0}=a$ (in
a relative sense) as $n$ increases is the substance of the argument
for the connection between intermittency and complex singularities.
In this setting, it is quite clear that we can add a polynomial or
any entire function to $f(t)$ without moving the singularity. Such
an addition can destroy intermittency in the coefficients of $(t-t_{0})^{n}$
for small $n$, but the intermittency will show up for larger $n$.
Thus intermittency may be identified more clearly in the signal $f(t)$
if the coefficients of $(t-t_{0})^{n}$ for certain small $n$ are
filtered out. In the numerical method we are describing, the range
of filtering is determined by examining the coefficients of the Chebyshev
series.

The right hand plot in Figure \ref{fig:lrz-filtering} shows a linear
fit of the filtered Chebyshev coefficients. The linear fit is obtained
as follows. An index $i$ is selected if and only if it is in the
range that survives filtering ($10<i<55$ in Figure \ref{fig:lrz-filtering})
and $|c_{i}|>|c_{j}|$ for $j>i$. The linear fit of $\log|c_{i}|$
vs $i$ is computed using only the selected indices $i$. The linear
fit approximates the outer envelope of $\log|c_{i}|$ vs $i$ as shown
in Figure \ref{fig:lrz-filtering}. If the slope of the fit is $\rho$,
\begin{equation}
\Re(t)=\frac{1}{2}\left(\rho+\frac{1}{\rho}\right)\cos\theta,\:\:\Im(t)=\frac{1}{2}\left(\rho-\frac{1}{\rho}\right)\sin\theta\label{eq:lrz-ellipse}
\end{equation}
is an ellipse in the complex plane parametrized by $\theta$. Standard
results in complex function theory imply that there must be a singularity
on this ellipse. The Padé approximant is far more reliable inside
the ellipse than outside it, as we will momentarily see.

\subsection{Computation of Padé approximant}

Given a Chebyshev series, its Padé approximant 
\[
\sum_{i=0}^{n}c_{i}T_{i}(t)\approx\frac{a_{0}+a_{1}T_{1}(t)+\cdots+a_{p}T_{p}(t)}{b_{0}+b_{1}T_{1}+\cdots+b_{q}T_{q}(t)}
\]
is computed using 
\[
\left(c_{0}+c_{1}T_{1}+\cdots+c_{n}T_{n}\right)\left(b_{0}+b_{1}T_{1}+\cdots+b_{q}T_{q}\right)-\left(a_{0}+a_{1}T_{1}+\cdots+a_{p}T_{p}\right)\approx0.
\]
If Chebyshev polynomials are multiplied using the identity $T_{i}T_{j}=\frac{1}{2}\left(T_{i+j}+T_{|i-j|}\right)$
and coefficients are equated to zero, we get a set of $n+1$ linear
relations between the unknown coefficients $a_{i}$ and $b_{i}$. 

We follow Gonnet, Pachón, and Trefethen \cite{GonnetPachonTrefethen2011}
in using the singular value decomposition of the last $(n-p)$ of
the equations to solve for the $b_{i}$ with the normalization $\sum|b_{i}|^{2}=1$.
The same authors recommend strategies to reduce the degree of the
dominator if the singular values are too small or if the smallest
singular values are repeated. We have experimented with these strategies,
but rely on different criteria to obtain robust Padé approximants.
Removing small singular values was problematic and threw away too
much information. In fact, in most of our computations we prefer to
have denominators of large degree with the degree being nearly $200$
in some cases.

\begin{figure}

\centering{}\includegraphics[scale=0.4]{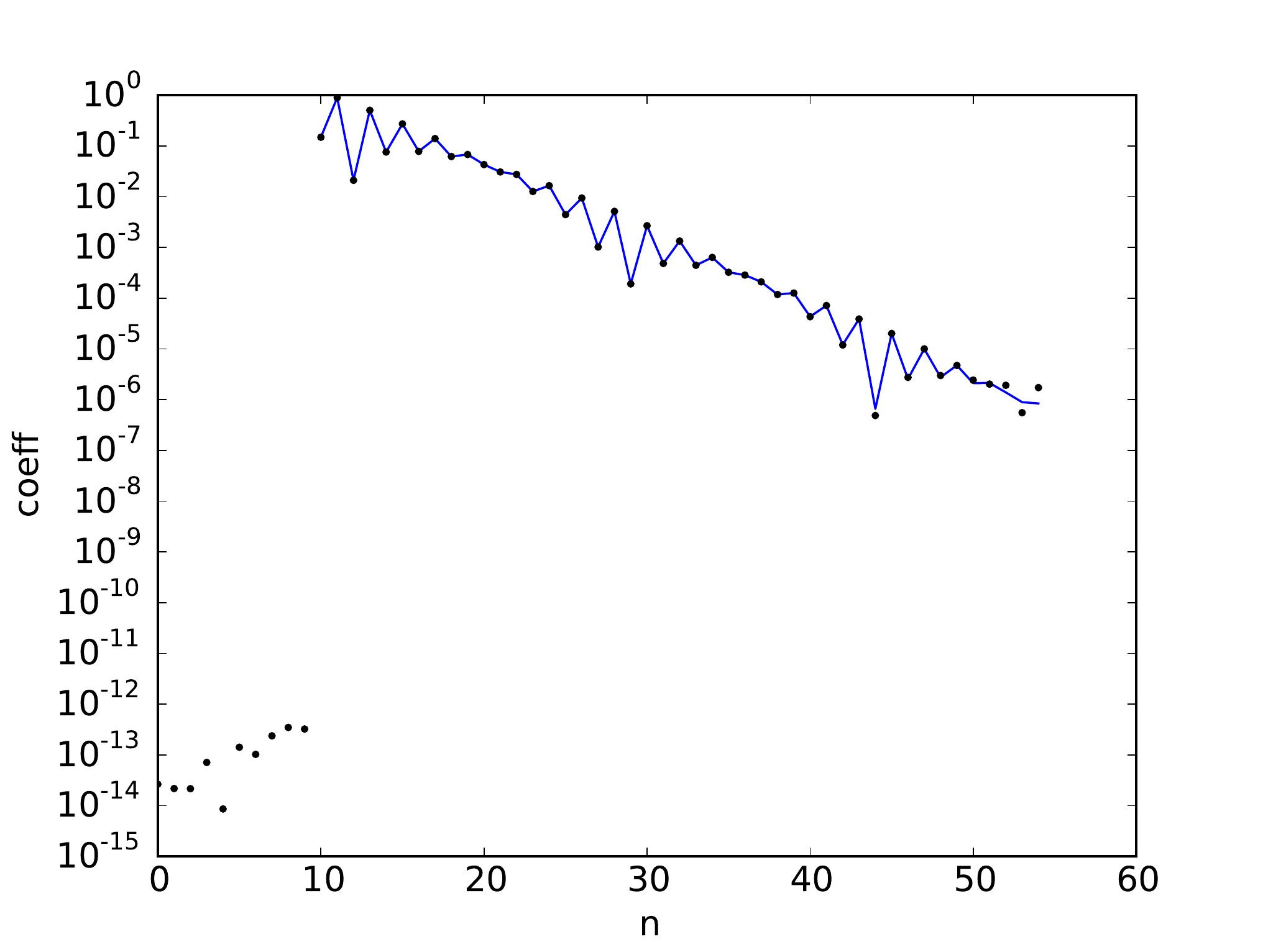}\caption{Comparison of filtered coefficients of Figure \ref{fig:lrz-filtering}
with coefficients of the Padé approximation. The Padé approximations
are dotted.\label{fig:lrz-fit-quality}}
\end{figure}

Figure \ref{fig:lrz-fit-quality} compares the filtered coefficients
and the Chebyshev coefficients of the Padé approximation. The Padé
approximation was obtained using $p=15$ and $q=25$. Since the first
$10$ coefficients are filtered out, the first $10$ coefficients
in the numerator are $0$. Most of the fitting is taking place in
the denominator. Close observation shows that the Padé approximation
gets the pattern near the tail slightly differently. However, the
quality of the approximation is excellent through the range of the
Chebyshev coefficients. Obtaining a high quality approximation of
this sort is essential. 

If two analytic functions agree exactly for $t\in[a,b]$, they will
agree everywhere in the complex plane. If one of those two functions
is a polynomial, it is incapable of picking up singularities of the
other function, even if the agreement over $t\in[a,b]$ is very close,
because polynomials are entire functions. The hope with Padé approximations
is that they will pick up the singularity in the complex plane because
they are rational in $t$. For that to be true, we need close agreement
over $t\in[a,b]$ at a minimum. In fact, such close agreement also
appears to be sufficient if the original signal is properly filtered.

\subsection{Phase plots}

As mentioned above, the left side plot in Figure \ref{fig:lrz-filtering}
gives a good indication of $\Re(t_{0})$ if $t_{0}$ is the location
of the singularity. The imaginary part $\Im(t_{0})$ may be obtained
by plotting the Padé approximant for complex values of $t$. Phase
plots \cite{Wegert2012,WegertSemmler2010} of a complex function $f(z)$
depict the function by indicating the phase $\arg f(z)$ using a color
code. Such plots reveal a wealth of information about poles and zeros.
According to Wegert and Semmler \cite{WegertSemmler2010}, the precise
origin of the idea of phase plots is difficult to determine. 

\begin{figure}

\centering{}\includegraphics[scale=0.4]{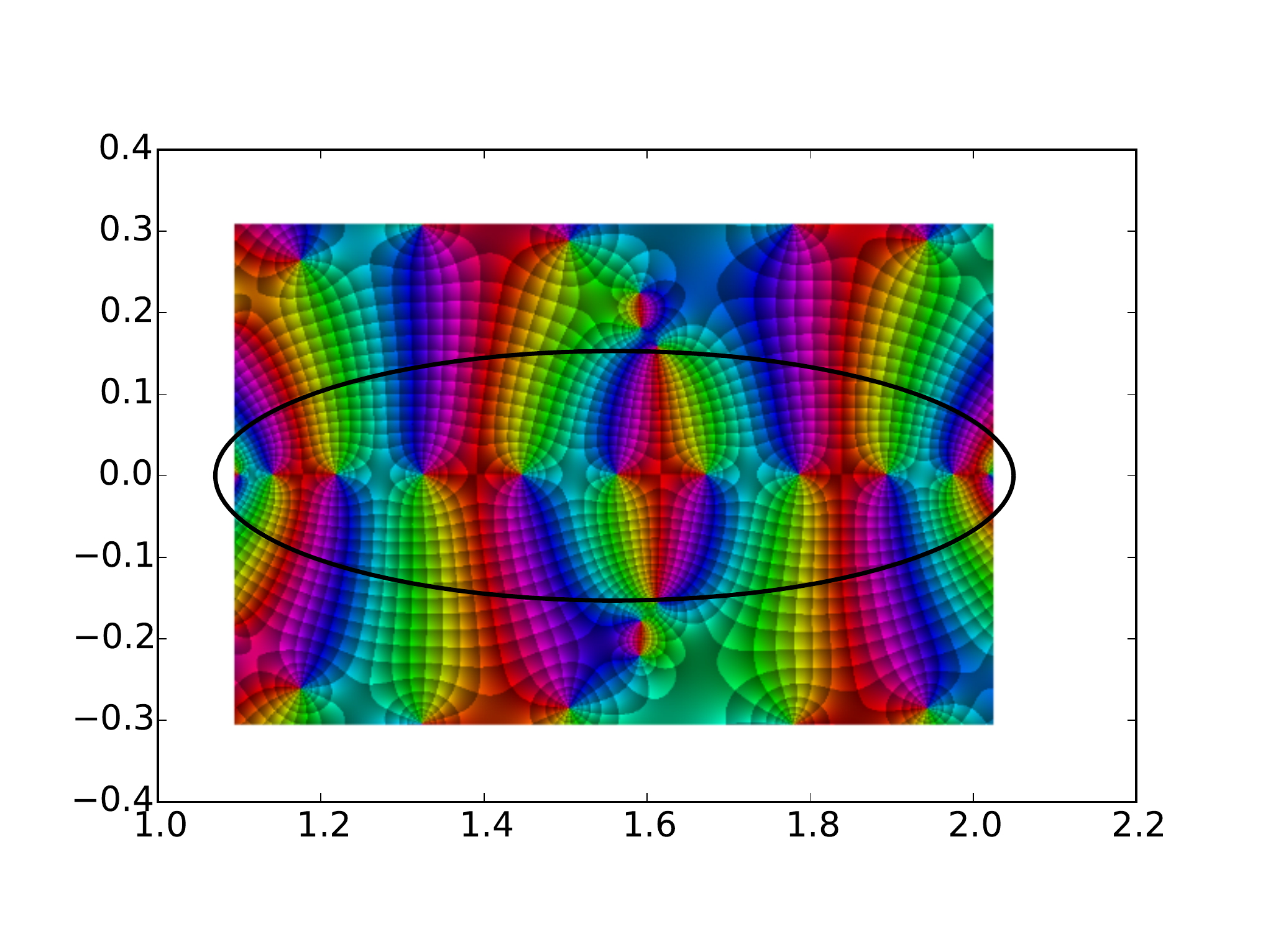}\includegraphics[scale=0.4]{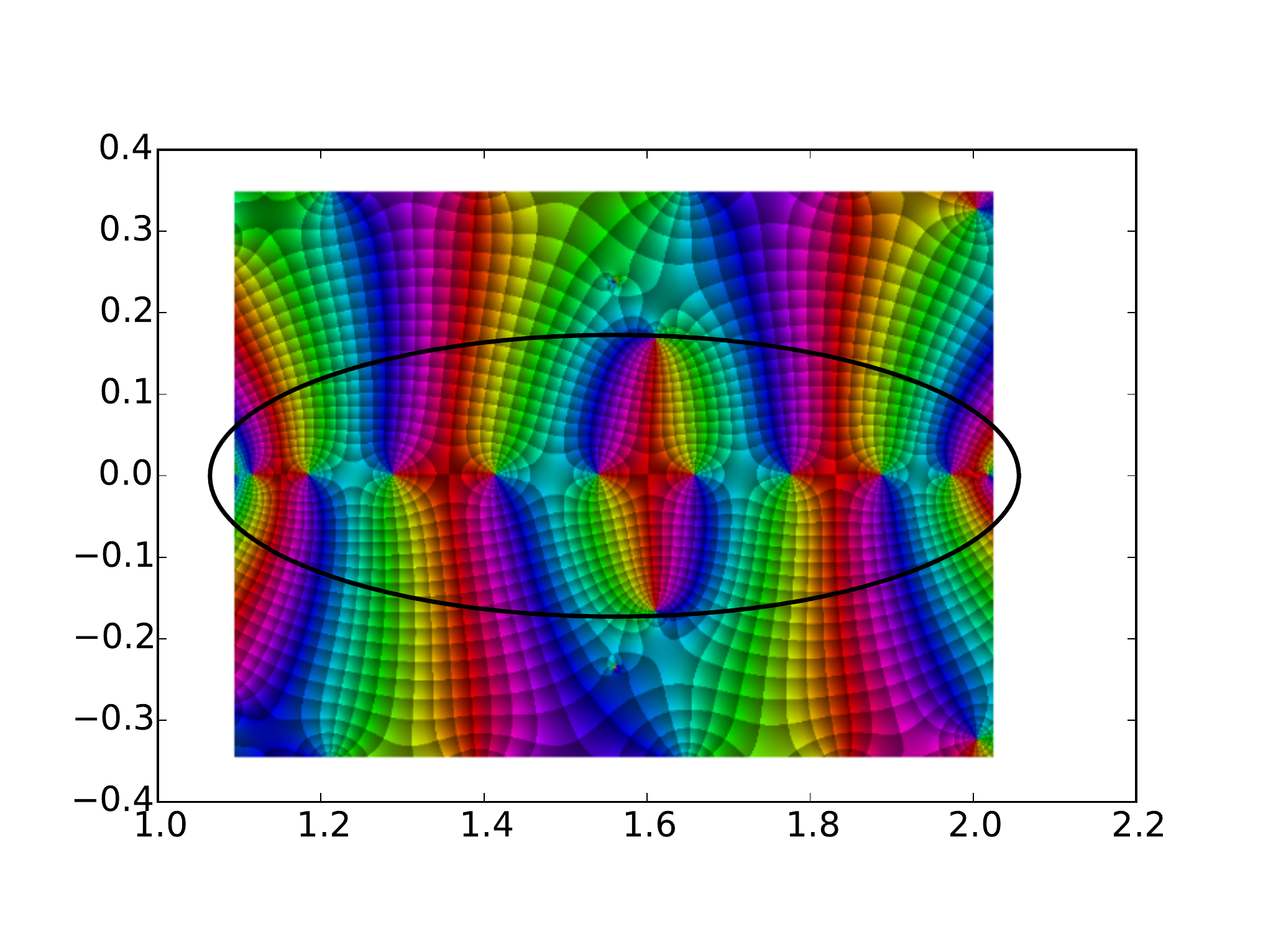}\caption{Phase plots of Padé approximants of the $z$ and $x$ coordinates,
respectively, of the Lorenz orbit $AABB$ corresponding to the middle
segment in Figure \ref{fig:lrz-seg-plots}. The colors code for the
phase or argument at a point in the complex $t$-plane. \label{fig:lrz-phase-plots}}
\end{figure}

Figure \ref{fig:lrz-phase-plots} shows phase plots for the $z$ (left
side) and $x$ (right side) coordinates of the middle segment of the
Lorenz orbit $AABB$. Each of the $11$ zeros of the filtered signal
shown in Figure \ref{fig:lrz-filtering} may be easily located in
the phase plot on the left. At a zero the colors red-blue-green rotate
in a counter-clockwise sense (opposite of Wegert's convention). The
ellipses shown as thick lines correspond to (\ref{eq:lrz-ellipse}). 

The phase plot also picks up a singularity near the boundary of the
ellipse and with $\Re(t)\approx1.6$ in both the plots. In the left
hand plot, corresponding to $z(t)$, the Padé approximant appears
to be attempting to pick up a double pole. The colors outside the
ellipse are not as reliable as those inside. Yet one may see that
the Padé approximant is attempting two clockwise rotations of the
colors red-blue-green. In the right hand plot, corresponding to $x(t)$,
there is clearly a single pole at the same location. The double pole
in the $z(t)$ phase plot is attempting to pick up the dominant part
in the psi-series (\ref{eq:lrz-psi-series}), which is a double pole
for $z(t)$. Similarly, the phase plot for $x(t)$ is picking a single
pole because the dominant part in the psi-series is a single pole.
In both cases, the artifacts of Padé approximation are well outside
the ellipse. 

\begin{table}
\begin{centering}
\begin{tabular}{|c|c|c|c|c|}
\hline 
Orbit & $x$ & $y$ & $z$ & accurate\tabularnewline
\hline 
\hline 
$AB$ & $0.175$ & $0.176$ & $0.172$ & $0.1714501006$\tabularnewline
\hline 
$AAB$ & $0.170$ & $0.167$ & $0.169$ & $0.1617621257$\tabularnewline
\hline 
$AAAB$ & $0.162$ & $0.158$ & $0.150$ & $0.1563426260$\tabularnewline
\hline 
$AABB$ & $0.170$ & $0.170$ & $0.156$ & $0.1636066901$\tabularnewline
\hline 
\end{tabular}
\par\end{centering}

\caption{Imaginary parts of the singularities closest to the real $t$-axis
for four Lorenz orbits. The $x,y,z$ columns are from singularity
locations determined using the $x,y,z$ coordinates, respectively.
The final column is from extended precision computations \cite{ViswanathSahutoglu2010}.\label{tab:lrz}}
\end{table}

Table \ref{tab:lrz} furthers validates the numerical method by comparing
the complex singularity location determined using $x$, $y$, and
$z$ coordinates with a more accurate computation from earlier work.
It appears that the singularity location is determined correctly with
about two digits of precision. Since the method for determining the
singularity is graphical, and based on mouse-clicks, two digits of
precision is about as much as we expect.

We have avoided finding roots of the polynomial in the denominator
of the Padé approximant. Polynomial root finding is often numerically
unsound. For the Lorenz orbits, denominator degrees range from $25$
to $40$. For the Couette orbits that we next turn to, the denominator
degrees can be larger than $100$ and polynomial root-finding is infeasible.
The phase plots are more reliable and informative.

\section{Application to plane Couette flow}

\begin{table}

\begin{centering}
\begin{tabular}{|c|c|c|c|c|c|c|c|}
\hline 
Orbit & $Re_{\tau}$ & $L$ & $M$ & $N$ & $dx^{+}$ & $dy^{+}$ & $dz^{+}$\tabularnewline
\hline 
\hline 
P1 & $28$ & $32$ & $64$ & $48$ & $4.9$ & $1.4$ & $2.2$\tabularnewline
\hline 
P2 & $33$ & $48$ & $64$ & $48$ & $3.8$ & $1.6$ & $2.6$\tabularnewline
\hline 
P3 & $33$ & $64$ & $56$ & $48$ & $2.9$ & $1.9$ & $2.6$\tabularnewline
\hline 
P4 & $33$ & $48$ & $64$ & $48$ & $3.8$ & $1.6$ & $2.6$\tabularnewline
\hline 
P5 & $34$ & $48$ & $64$ & $48$ & $3.9$ & $1.7$ & $2.7$\tabularnewline
\hline 
P6 & $34$ & $48$ & $64$ & $48$ & $3.9$ & $1.7$ & $2.7$\tabularnewline
\hline 
\end{tabular}\caption{Grid resolution data for $6$ periodic or relative periodic solutions
of plane Couette flow.\label{tab:couette-grid-resolution}}

\par\end{centering}

\end{table}

The velocity field of plane Couette flow is represented as

\[
{\bf u}(x,y,z,t)=\sum_{\ell=-L/2}^{L/2}\sum_{n=-N/2}^{N/2}\sum_{m=0}^{M}{\bf u}_{l,m,n}(t)\exp\left(\frac{i\ell x}{\Lambda_{x}}+\frac{inz}{\Lambda_{z}}\right)T_{m}(y).
\]
Thus the velocity field is periodic in $x$ and $z$ with periods
equal to $2\pi\Lambda_{x}$ and $2\pi\Lambda_{z}$. For the minimal
flow unit, these periods are $5.4977..$ and $3.7699...$, respectively
\cite{HamiltonKimWaleffe1995}. As before, $T_{m}$ are Chebyshev
polynomials. At the walls $y=\pm1$, the velocity field satisfies
the boundary condition ${\bf u}=(\pm1,0,0)$. For the minimal flow
unit, the Reynolds number in the incompressible Navier-Stokes equations
\[
\frac{\partial{\bf u}}{\partial t}+({\bf u}.\nabla){\bf u}=-\nabla p+\frac{1}{Re}\triangle{\bf u}
\]
is $Re=400$. The pressure $p$ is determined by the incompressibility
constraint $\nabla.{\bf u}=0$.

Table \ref{tab:couette-grid-resolution} shows resolution data for
six periodic or relative periodic solutions of the minimal flow unit.
A more detailed description of these orbits may be found in \cite{Viswanath2007}.
The orbits are labeled P1 through P6. P1 was first computed by Kawahara
and Kida \cite{KawaharaKida2001}. 

The frictional Reynolds number $Re_{\tau}$ of all six orbits is low,
although they exhibit some features of turbulence such as a sharp
peak in turbulence energy production in the near wall region. The
orbits capture other aspects of buffer layer dynamics as well \cite{Viswanath2007}.
The purpose of the minimal flow unit is to capture certain essential
features of turbulence in as small a system as possible. The fact
that the Reynolds number is low, makes it easier to reach into the
high wave-number region where intermittency manifests itself most
clearly.

Table \ref{tab:couette-grid-resolution} gives the spacing between
grid points in the $x,y,z$ direction in frictional units. For the
non-uniform Chebyshev grid in the $y$-direction, the reported number
is the maximum spacing. There is no question that all six orbits are
exceedingly well-resolved. Good resolution is essential for reaching
wave-numbers where intermittency is pronounced.

The Lorenz system is only $3$ dimensional. For the minimal flow unit,
the dimensionality of the spatial discretizations we use is more than
$10^{5}$. Since the numerical method of the previous section is visual,
only a small fraction of these modes can be examined. We examine about
$50$ modes for each orbit. These modes are indexed by $k$. Let $\delta\ell=L/64$,
$\delta m=M/32$, and $\delta n=N/64$. For $0\leq k\mod16<6$ and
$k=15$, these modes are $\ell,\, m,\, n\approx1+j\delta\ell,\,1+j\delta m,\,1+j\delta n$,
where $j=k\mod16$. For $6\leq k\mod16<12$, we use $\ell=m=n=j-5$,
where $j=k\mod16$ as before. For $12\leq k\mod16<15$, we use $\ell,\, m,\, n=L/4+(j-14),\, M/2+(j-14),\, N/4+(j-14)$,
where $j=k\mod16$ as before.

If ${\bf u}=(u,v,w)$, we extract the stream-wise $u$ mode if $0\leq k<16$,
the wall-normal $v$ mode if $16\leq k<32$, and the span-wise $w$
mode if $32\leq k<48$. Most of these modes are complex valued, and
we retain only the real part in the signal that is extracted. 

The selection of modes for $0\leq k<48$ is admittedly a little arbitrary.
The aim is to look at a few low wave-numbers, a few in the middle,
and a few higher wave-numbers. The choice of wave-numbers proved adequate
for demonstrating the connection between intermittency and complex
singularities in all six orbits.

\begin{table}

\centering{}%
\begin{tabular}{|c|c|c|c|c|c|c|}
\hline 
Orbit & $k$ & $n_{fst}$ & $n_{lst}$ & $p$ & $q$ & $d$\tabularnewline
\hline 
\hline 
P1 & $19$ & $12$ & $65$ & $18$ & $45$ & $12.0$\tabularnewline
\hline 
P2 & $24$ & $50$ & $200$ & $98$ & $100$ & $4.3$\tabularnewline
\hline 
P3 & $7$ & $45$ & $140$ & $68$ & $70$ & $5.5$\tabularnewline
\hline 
P4 & $0$ & $200$ & $400$ & $218$ & $180$ & $2.6$\tabularnewline
\hline 
P5 & $10$ & $75$ & $250$ & $98$ & $150$ & $3.8$\tabularnewline
\hline 
P6 & $1$ & $100$ & $250$ & $123$ & $125$ & $4.3$\tabularnewline
\hline 
\end{tabular}\caption{Data for singularity computations. The $k$ column gives the mode
for which the Chebyshev series in time had the best pattern. The interpretation
of $k$ is found in the text. All Chebyshev modes $c_{i}$ outside
of $(n_{fst},n_{lst}]$ were filtered out. The $p$ and $q$ columns
give the degrees of the Padé numerator and denominator, respectively.
Notice that most of the coefficients in the numerator are nearly zero
because of filtering. The last column gives the estimated distance
of the singularity from the real $t$-axis.\label{tab:couette-singularity-data}}
\end{table}

\begin{figure}

\begin{centering}
\includegraphics[scale=0.3]{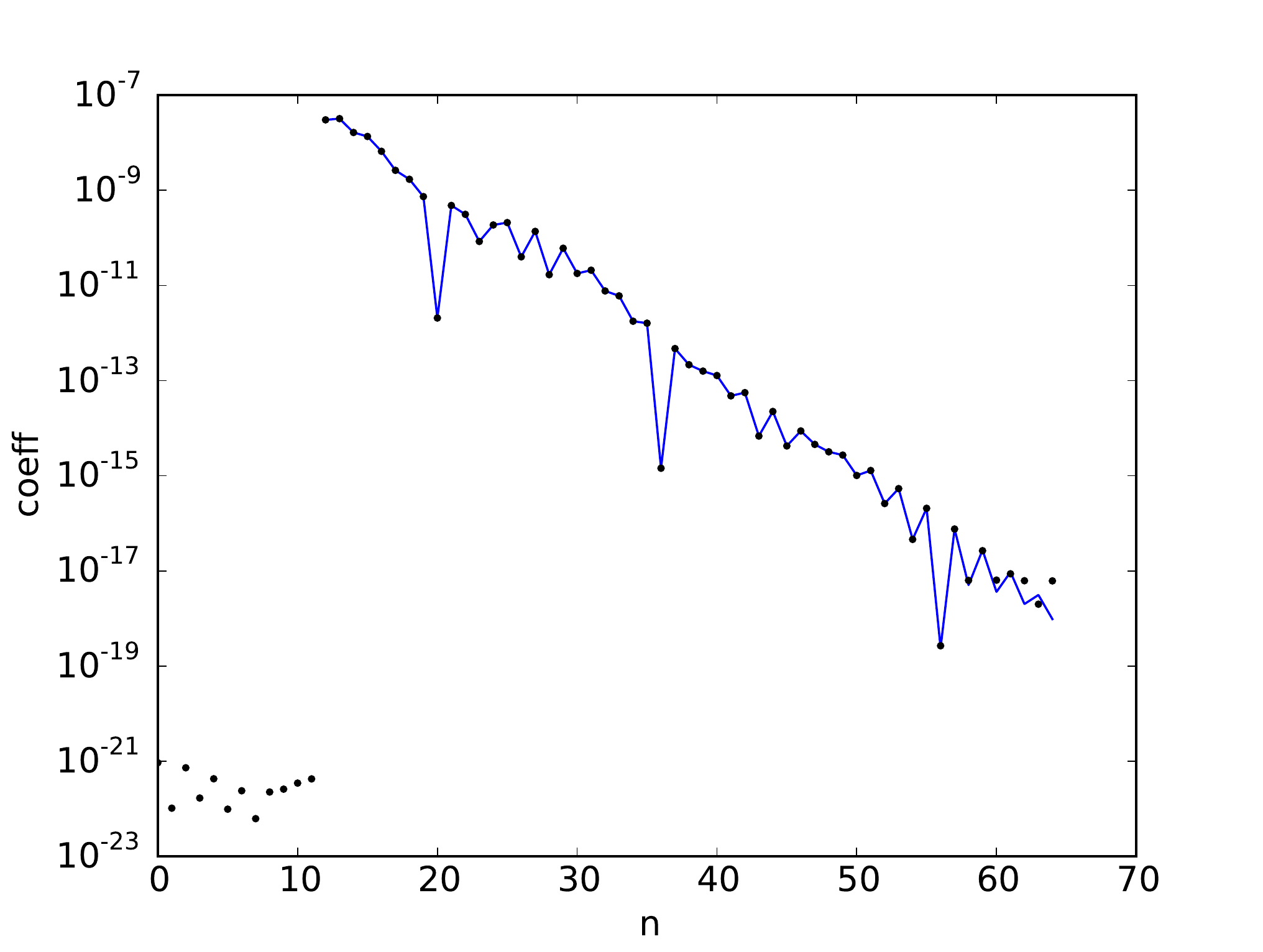}\includegraphics[scale=0.3]{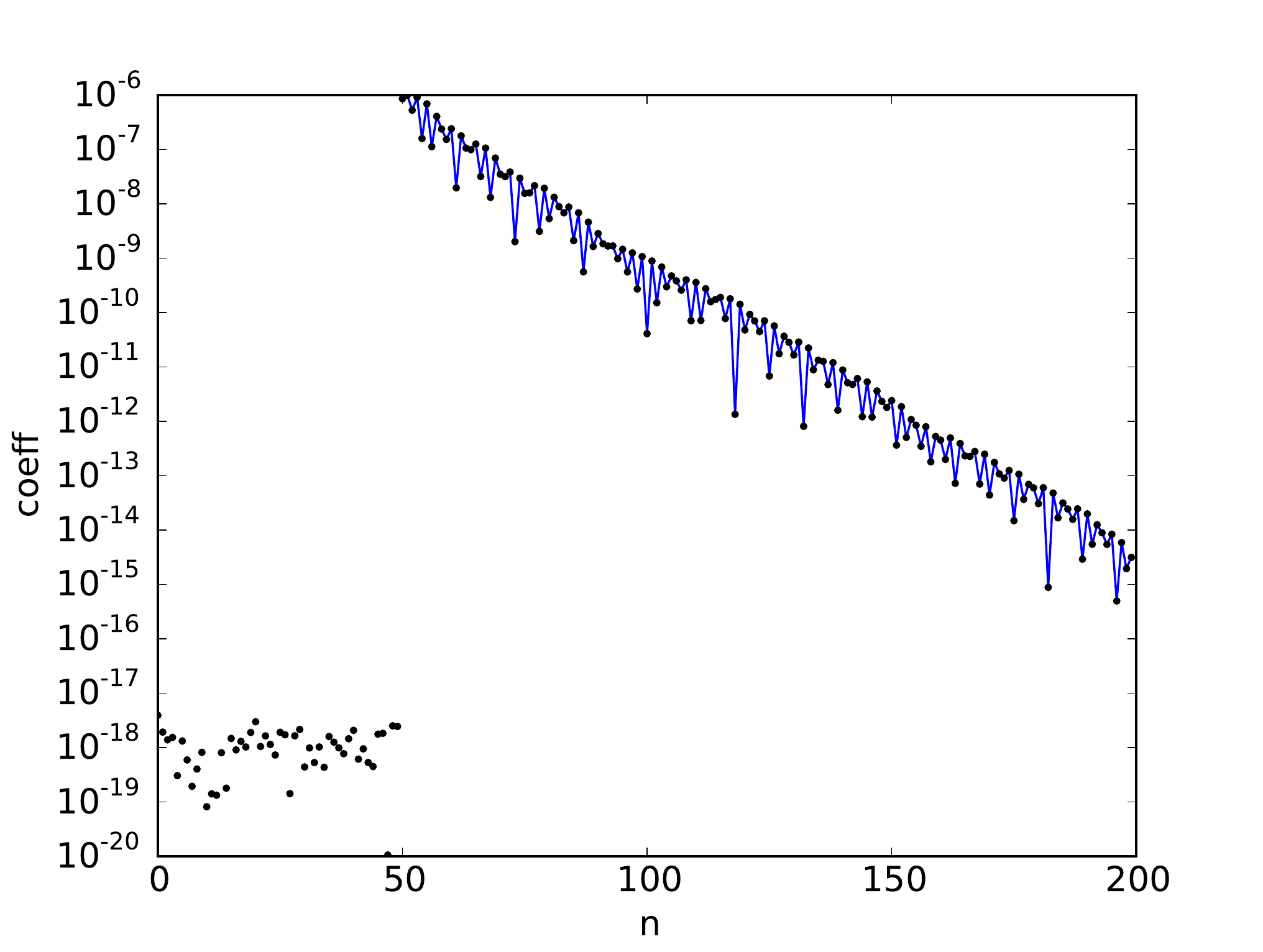}
\par\end{centering}

\centering{}\includegraphics[scale=0.3]{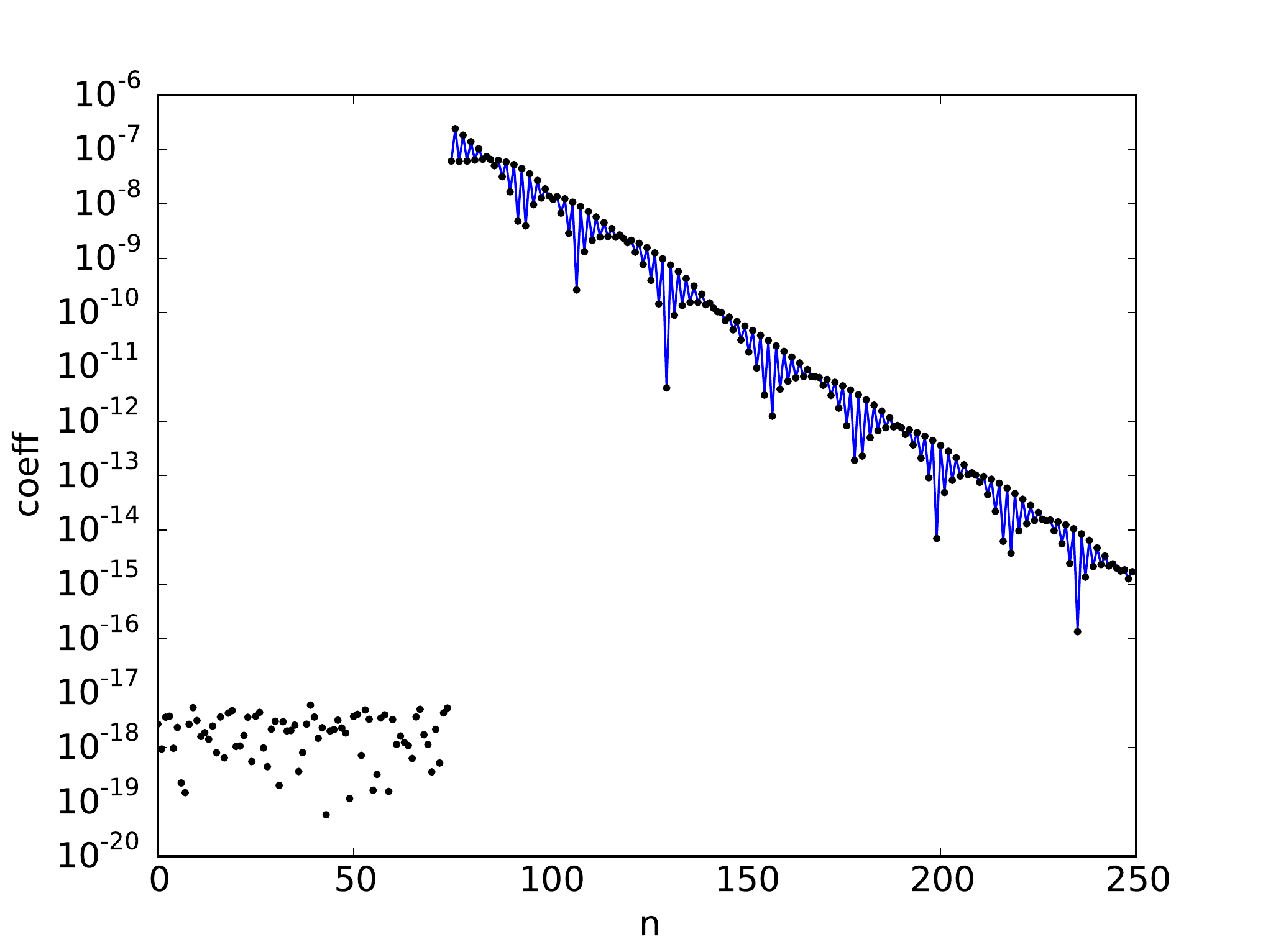}\includegraphics[scale=0.3]{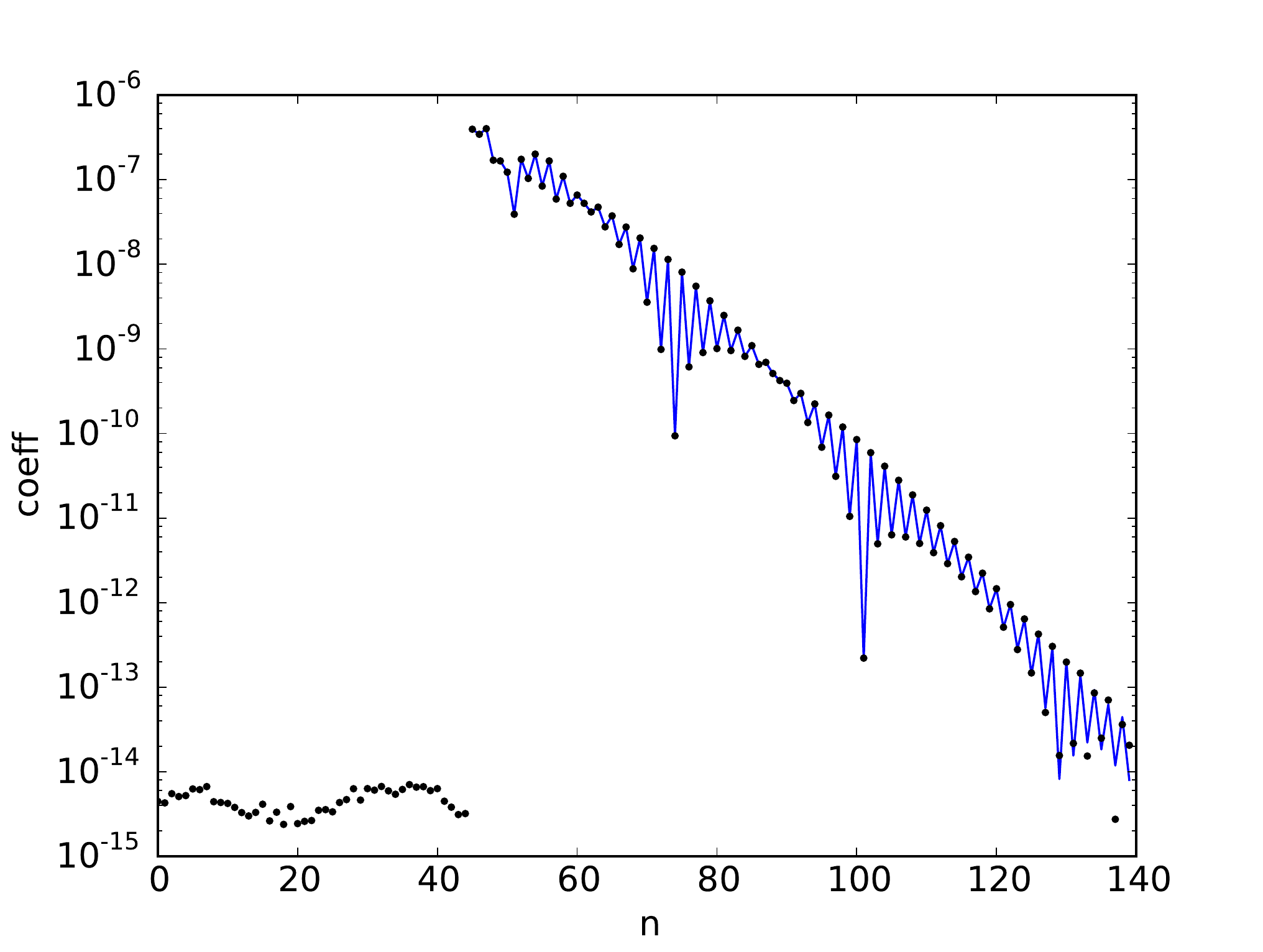}\caption{The filtered Chebyshev series for P1, P2, P3, and P5 (clockwise from
top-left). The $y$-axis is the absolute value of the Chebyshev coefficient
of $T_{n}(t)$. The circular markers are approximations computed using
the Padé approximant.\label{fig:couette-fit-quality}}
\end{figure}

Table \ref{tab:couette-singularity-data} shows the parameters used
for computing the complex singularities. The mode $k$ and the filtering
interval are chosen so that the coefficients of Chebyshev series in
time exhibit a clean pattern. Plots of the filtered series are shown
in Figure \ref{fig:couette-fit-quality} for P1, P2, P3, and P5. In
each case, a clear pattern is discernible, and the Padé approximants
reproduce the coefficients with excellent accuracy. 

The plots in Figure \ref{fig:couette-fit-quality} pick modes corresponding
to values of $k$ listed in Table \ref{tab:couette-singularity-data}.
We tried to pick the $k$ for which the pattern in the Chebyshev series
was the cleanest. Although there are other choices of $k$ that are
nearly as good, a pattern shows up cleanly only for a few values of
$k$.

The need to select $k$ carefully may be explained as follows. In
complex function theory, $f(z)$ is usually a complex valued function
of a complex variable. However, much of complex function theory goes
through with little change even if $f(z)$ takes values in a Hilbert
space or a Banach space, as long as $z$ is a complex variable. For
instance, Cauchy's theorem continues to be true. 

We may think of ${\bf u}(t)$, the velocity field, as an analytic
function of $t$ which takes values in some Hilbert space, although
the velocity field is computed only for real values of $t$. The singularity
at a point $t_{0}$ in the complex $t$-plane may be thought of as
a series of some type in $(t-t_{0})$, something like the psi-series
in (\ref{eq:lrz-psi-series}), with coefficients in a Hilbert space.

When we pick a mode corresponding to some $k$ and record the variation
of that mode in time, we are essentially taking an inner product of
the mode and the expansion of the singularity. The Padé approximant
tries to pick up the dominant part of the singularity. Two random
vectors in high dimensional space are nearly orthogonal. Ordinarily
the dominant part of the singularity does not project well when an
inner-product is taken. Thus the need to pick $k$ carefully.

\subsection{Intermittency and singularities in the complex $t$-plane}

\begin{figure}

\begin{centering}
\includegraphics[scale=0.4]{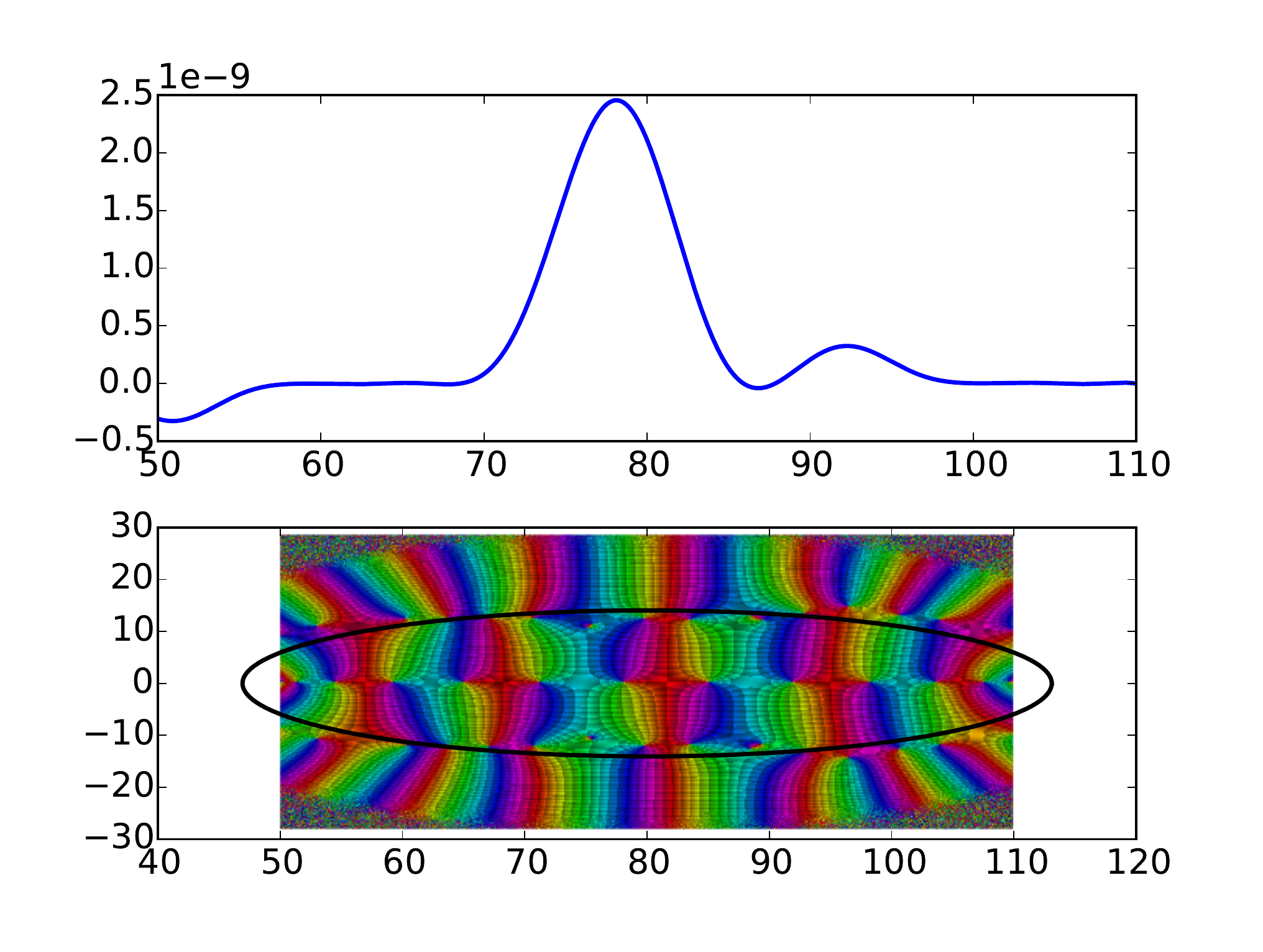}\includegraphics[scale=0.4]{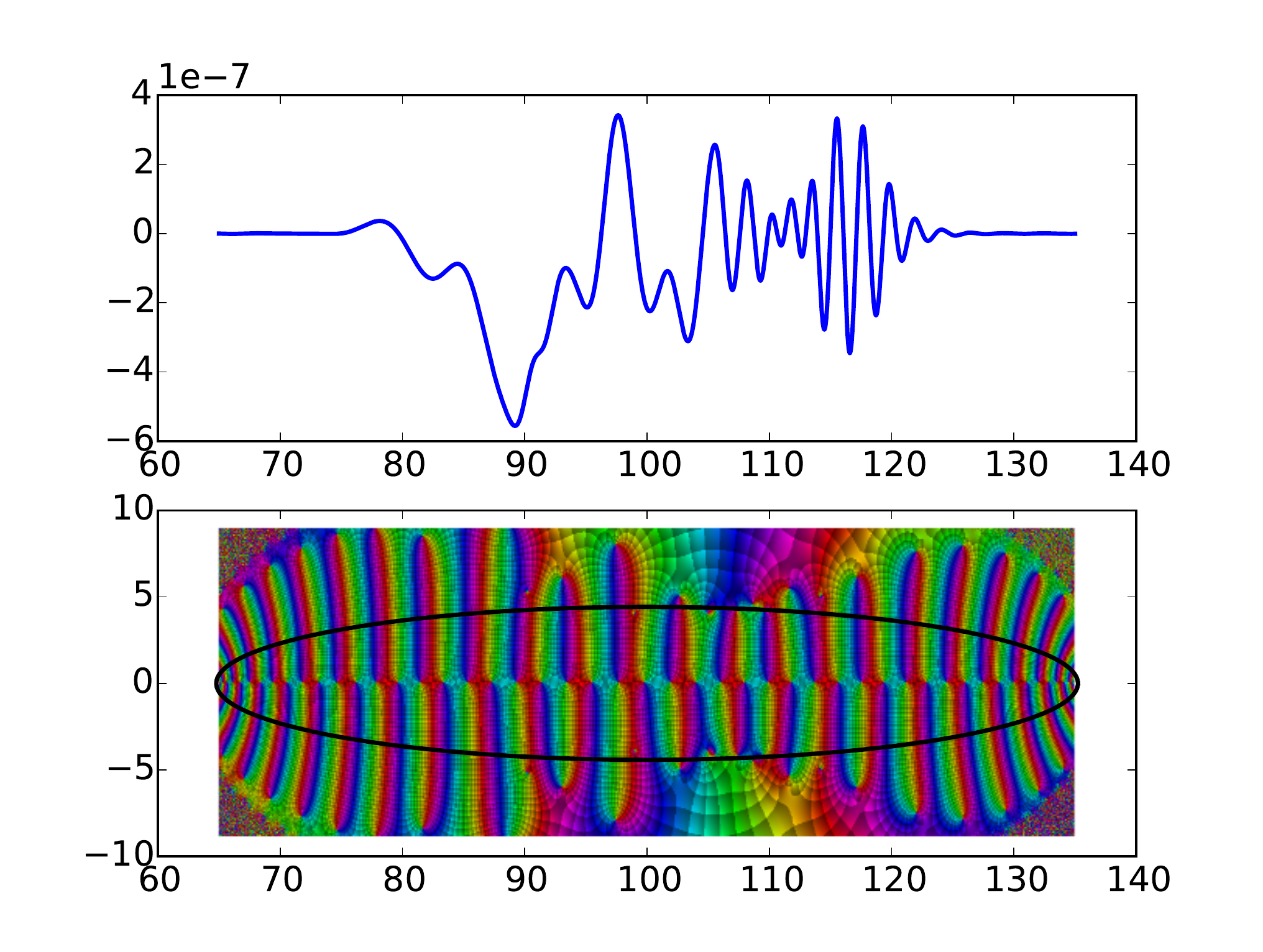}
\par\end{centering}

\centering{}\includegraphics[scale=0.4]{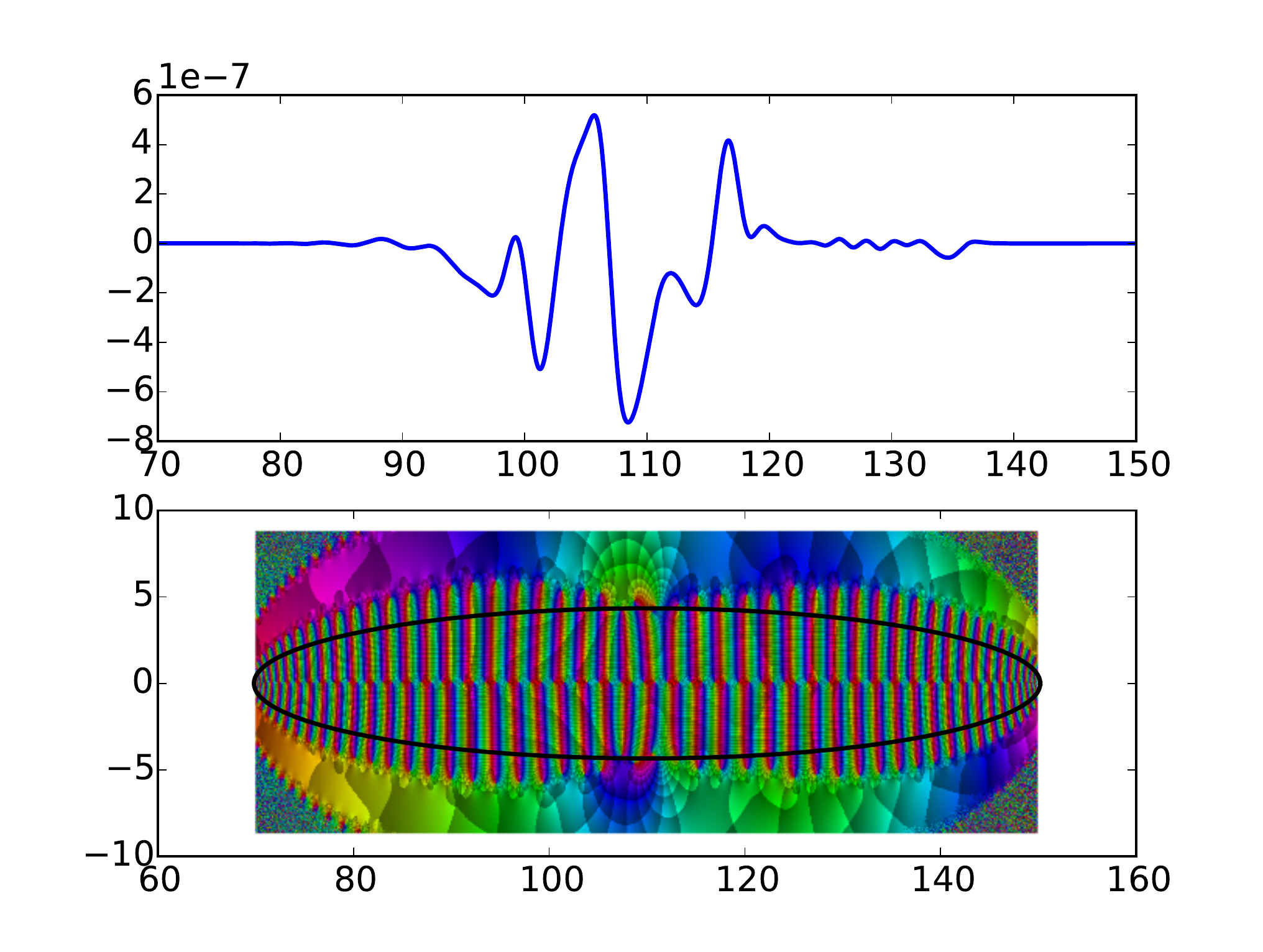}\includegraphics[scale=0.4]{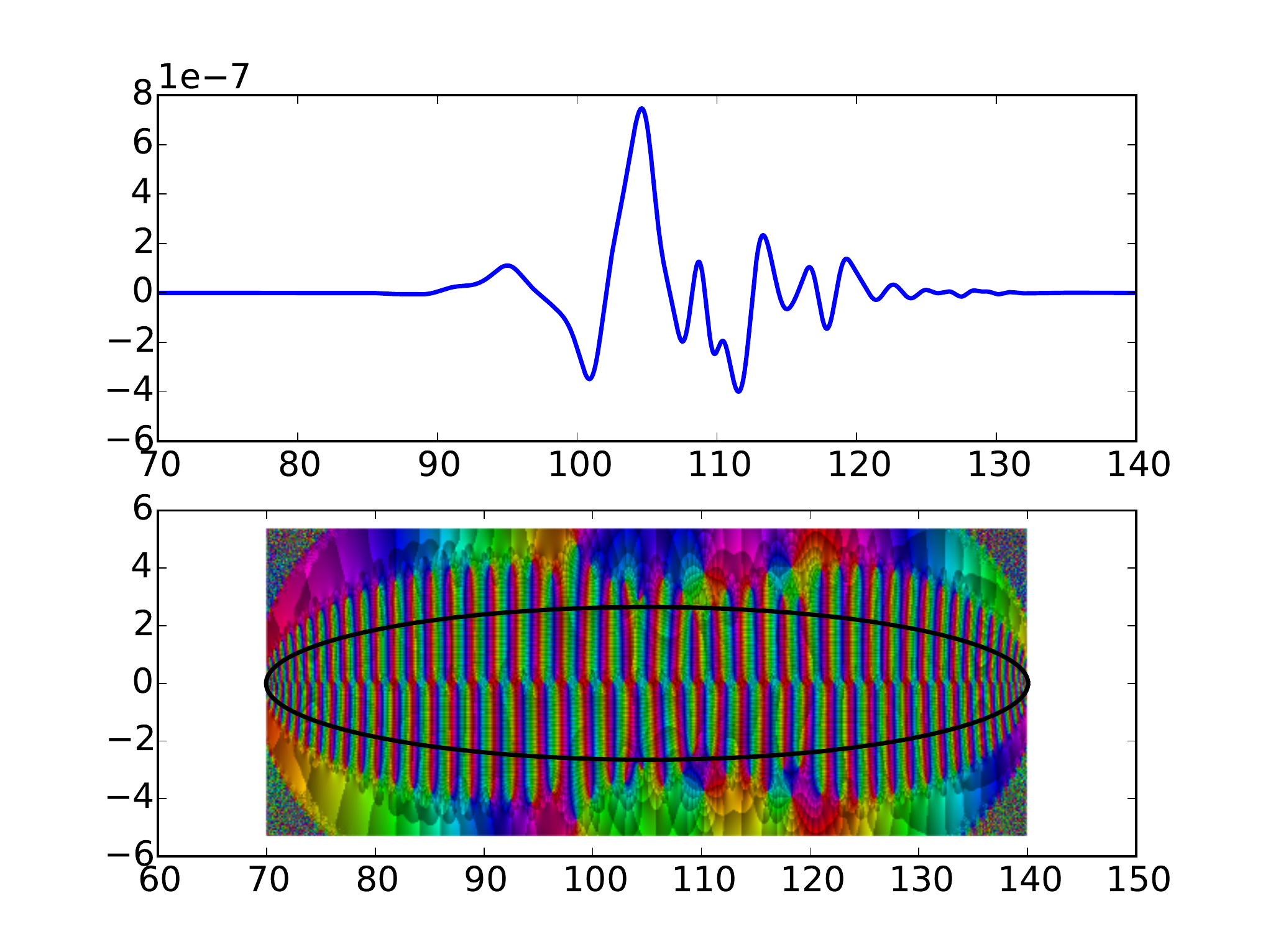}\caption{An intermittent high-wavenumber signal as a function of time and a
phase plot of the Padé approximant after filtering for some other
mode (the mode corresponding to $k$ listed in Table \ref{tab:couette-singularity-data})
for the orbits P1, P2, P4, and P6 (clockwise from top left).\label{fig:couette-intermittency}}
\end{figure}
The connection between complex singularities and intermittency for
solutions of the minimal flow unit is demonstrated in Figure \ref{fig:couette-intermittency}.
For each of the four orbits shown, the top plot is an intermittent
high wave-number signal that was not filtered. The mode that was chosen
corresponds to $k$ with $12\leq k\mod16<16$, with the interpretation
of $k$ given above. Almost any $k$ in this range can be picked.

All the phase plots, except for P1, are clean inside the ellipse.
For P4 and P5, there is a singularity near the boundary of the ellipse,
corresponding to the intermittency, but no spurious singularities
are found inside the ellipse. The intermittency in P2 is not as pronounced
and there may be more one singularity near the border of the ellipse.
The filtered signals corresponding to the phase plots are not shown.
Yet the oscillation of these signals with $t$ for real $t$ is clear
from the phase plots. When interpreting the phase plots, it must be
borne in mind that a counter-clockwise cycle of red-blue-green is
a zero, while a clockwise cycle is a pole%
\footnote{This is the opposite of Wegert's convention.%
}.

Following \cite{FrischMorf1981,Kraichnan1967}, the reason that intermittency
manifests itself in the high wave-numbers, without filtering, but
is more pronounced only after filtering in the lower wave-numbers
may be explained as follows. The energy in the spatial modes of the
velocity field decreases exponentially with wave-number in the dissipative
range. At a given point in time, with $t$ real, this rate is related
in some way to the distance to the nearest singularity in the complex
$t$-plane. The rate appears to be highest right below the singularity.
A small change in the rate causes a much bigger change in the relative
magnitude of the energy in the high wave-numbers, making intermittency
quite pronounced as evident from Figure \ref{fig:couette-intermittency}.

\subsection{Distance of the singularities from the real axis}

We do not claim much accuracy for the distances of the singularities
from the real $t$-axis reported in Table \ref{tab:couette-singularity-data}.
The errors may be as high as $20$\%. Nevertheless, it is very clear
that P1's singularity is much farther way than that of P2 through
P6.

Within the confines of the minimal flow unit, P2 through P6 are all
in the turbulent regime. However, P1 is intermediate between laminar
flow and turbulent flow \cite{KawaharaKida2001,Viswanath2007}. The
laminar solution of course is constant in time and therefore an entire
function. P1's singularity seems to be much farther out because it
is intermediate between laminar flow and turbulent flow.

Another question we may ask is about the type of the singularities.
Are they poles of some order, or are they approximable by a pole of
some order? For Lorenz, we know that the psi-series are convergent
singular solutions. Although not all singularities are proved to be
of that form, a few singularities have been analyzed in extended precision
and verified to be psi-series \cite{ViswanathSahutoglu2010}. The
dominant part in the psi-series is either a pole or a double pole.
Thus the success of Padé is not a big surprise.

In the case of the incompressible Navier-Stokes equations, the analytic
form of singular solutions is not known. The Padé approximants use
poles to approximate all kinds of singularities and do not give direct
information about the form of the singularities. Yet the clarity with
which Padé approximants locate the singularities, as shown by the
phase plots of Figure \ref{fig:couette-intermittency}, makes us think
that the form of the singularities is unlikely to be much more complex
than for Lorenz.

Another related question is about the number of singularities. Each
of the orbits P1 through P6 seems to be governed by a single dominant
singularity, or a tight cluster of singularities closest to the real
axis in the case of P2. For a long trajectory and in the complex $t$-plane
as a whole, it appears reasonable to suspect that singularities are
numerous and occur with increasing density as the Reynolds number
increases.

\section{Discussion}

Computations described in the previous section, even though limited
to the minimal flow unit of plane Couette turbulence, offer definite
evidence that intermittency in fine scales is due to singularities
in the complex plane as suspected by Frisch and others \cite{Frisch1995,FrischMorf1981}. 

All our computations of complex singularities are in the time domain.
However, intermittency also has a significant spatial aspect \cite{Sreenivasan1991}.
The connection between singularities in the time domain and spatial
aspects of intermittency is not entirely clear. It may be possible
to combine spatial modes in a manner that the temporal intermittency
becomes particularly pronounced.

On the other hand, the velocity field at a fixed value of time may
have singularities if the spatial coordinates are allowed to be complex.
It is unclear what such spatial singularities look like and how they
relate to temporal singularities.

All six orbits discussed in the previous section go through one cycle
of the self-sustaining process in a single period \cite{Viswanath2007}.
The self-sustaining process as developed by Waleffe \cite{Waleffe1997}
is characterized in the spatial domain in terms of rolls and streaks.
Each of the six orbits of the minimal flow unit is governed by a single
dominant singularity (and its complex conjugate), or perhaps by a
tight cluster of singularities. The possibility of a connection between
the self-sustaining process and the location of singularities in the
complex $t$-plane may be worth investigating.

The finite time blow-up problem for the incompressible Navier-Stokes
equations has been viewed in light of complex singularities \cite{BiswasFoias,FrischMatsumotoBec2003}.
The robustness with which complex singularities are located and visualized
in Figure \ref{fig:couette-intermittency} suggests that computing
the distance of the singularities from the real $t$-axis as a function
of the Reynolds number may shed some light on this topic. However,
there is a subtle difference between mathematical and computational
investigations of finite time blow-up.

The evidence that direct numerical simulation algorithms solve the
Navier-Stokes equations is overwhelming. Yet there is a subtle difference
between numerical and mathematical investigations. In mathematical
formulations, the velocity field of the Navier-Stokes equations is
allowed to belong to certain function spaces, and these function spaces
are much wider than the class of velocity fields that are subject
to numerical simulation. Any direct numerical simulation code will
become unstable if the initial velocity fields are not sufficiently
regular. At high Reynolds numbers such as $Re_{\tau}=1000$, establishing
a turbulent initial state can pose far greater difficulties than sustaining
turbulence. In fact, at such Reynolds numbers, attempting to transition
to turbulence by adding a small disturbance to laminar flow is nearly
impossible because of the large number of transitional instabilities.
A turbulent initial condition is established by slowly raising the
Reynolds number. Thus turbulent velocity fields are pre-selected in
a way to induce statistical steadiness in the flow with respect to
time integration.

\section{Acknowledgements}

We are very thankful to Nick Trefethen, and his students Hrothgar
and Marcus Webb, for sharing their work on Padé approximation with
us. This research was partially supported by NSF grant DMS-1115277. 

\bibliographystyle{plain}
\bibliography{references}

\end{document}